\documentclass[fleqn]{svjour2}
\smartqed  % flush right qed marks, e.g. at end of proof
\usepackage{graphicx}
\usepackage[version=3]{mhchem}
\usepackage[dvipsnames]{xcolor}
\usepackage{rotating}
\usepackage{mciteplus}
\usepackage{color,soul}
\usepackage[sort&compress,comma,square,super]{natbib}
\usepackage{mathptmx}      % use Times fonts if available on your TeX system
\newcommand{\cc}[1]{{\color{red}#1\color{black}}}
\newcommand{\E}{\widehat{E}}
\newcommand{\ee}{\mathrm{ee}}

\newcommand{\HF}{\mathrm{HF}}

\newcommand{\CC}{\mathrm{CC}}

\newcommand{\e}{\widehat{e}}
\newcommand{\ad}{\widehat{a}^\dagger}
\newcommand{\an}{\widehat{a}}

\newcommand{\ie}{{\it i.e.,\ }}
\newcommand{\eg}{\textit{e.g.\,}}
\journalname{Theoretical Chemistry Accounts}
\begin{document}

\title{Fermi and Coulomb correlation effects upon the interacting
quantum atoms energy partition}


\author{Isela Ruiz \and
        Eduard Matito \and 
        Fernando Jos\'e Holgu\'in-Gallego \and
        Evelio Francisco \and
        \'Angel Mart\'in Pend\'as \and
       Tom\'as Rocha-Rinza %etc.
}


\institute{I. Ruiz \and F. J. Holgu\'in-Gallego \and T. Rocha-Rinza \at
Institute of Chemistry, 
National Autonomous University of Mexico, Circuito Exterior, Ciudad
Universitaria, Delegaci{\'o}n Coyoac{\'a}n C.P. 04510, Mexico
City, Mexico.\\
              \email{tomasrocharinza@gmail.com}           
           \and
           E. Matito \at
Kimika Fakultatea, Euskal Herriko
Unibertsitatea (UPV/EHU) and Donostia International Physics Center (DIPC).
P.K. 1072, 20080 Donostia, Euskadi, Spain. 
IKERBASQUE, Basque Foundation for Science, 48011 Bilbao, Euskadi, Spain
           \and
           E. Francisco \and A. Mart\'in Pend\'as  \at
Department of Physical and Analytical
Chemistry, University of Oviedo, Juli{\'a}n Claver{\'i}a 8, Oviedo, Spain
}

\date{Received: April 20, 2016 / Accepted: }

\maketitle

\begin{abstract}

\noindent The Interacting Quantum Atoms (IQA) electronic energy partition is an
important method in the field of quantum chemical topology which has
given important insights of different systems and processes in
physical chemistry. 
%Recently, we have included Electron Correlation
%(EC) by means of coupled cluster theory using HF/CC transition density
%matrices (Ch{\'a}vez-Calvillo et al., \textit{Comput. Theor. Chem.},
%\textbf{2015}, \textit{1053}, 90).
There have been several attempts to include Electron Correlation (EC)
in the IQA approach, for example, through DFT and
Hartree-Fock/Coupled-Cluster (HF/CC) transition densities.
%$\varrho^{\mathrm{HF/CC}}_1(\mathbf{r}_1, \ \mathbf{r}_1^{\,
%\prime})$ and $\varrho^{\mathrm{HF/CC}}_2(\mathbf{r}_1, \
%\mathbf{r}_2)$. 
This work addresses the separation of EC in Fermi and
Coulomb correlation and its effect upon the IQA analysis by taking
into account spin-dependent one- and two-electron matrices 
$D^{\mathrm{HF/CC}}_{p\sigma q \sigma}$ and 
$d^{\mathrm{HF/CC}}_{p\sigma q\sigma r\tau s\tau}$ wherein $\sigma$ and
$\tau$ represent either of the $\alpha$ and $\beta$ spin projections.
We illustrate this approach by considering
\ce{BeH2}, \ce{BH}, \ce{CN-}, \ce{HF}, \ce{LiF}, \ce{NO+},
\ce{LiH}, \ce{H2O\bond{...}H2O} and \ce{HC#CH\bond},
which comprise non-polar 
covalent, polar covalent, ionic and hydrogen bonded systems.
The same and different spin contributions to
(\textit{i}) the net, interaction and exchange-correlation
IQA energy components
and (\textit{ii}) delocalisation indices defined in the quantum theory of
atoms in molecules are carefully examined and discussed.
 %In addition, we examined the spin-contributions of the 
%intra- and interactomic IQA energies in
%the open shell systems \ce{CH3.}, \ce{CH3CH2.}, \ce{(CH3)2CH.} and
%\ce{(CH3)3C.}.
Overall, we expect that this kind of analysis
will yield important insights about Fermi and Coulomb correlation in
covalent bonding, intermolecular interactions and electron
delocalisation in physical chemistry.

\end{abstract}

\section{Introduction}
\label{intro}

Wavefunctions analyses are aimed to get chemical insights from electronic
structure calculations. Unfortunately, there are many concepts in
chemistry \emph{e.g.} 
ar\-o\-ma\-tic\-ity, chemical bonds, electron delocalisation
and atomic charges which 
are not ob\-serv\-a\-bles.~\cite{frenking:07jcc}
%do not appear explicitly in the
%electronic Hamiltonian or the wave function $| \Psi \rangle$. 
Hence, there is not an unique way to compute quantities related with
such intuitive chemical notions. For example, there are 
 %This has been attempted for example through
orbital-based approaches such as Mulliken~\cite{mulliken:55jcp} 
and L{\"o}wdin~\cite{lowdin:50jcp} population
schemes which have been developed for the calculation of atomic
charges in molecular and supramolecular systems.
Nonetheless, these techniques have the disadvantage of being very
dependent on the particular elements used to built the wavefunction
like the basis set.~\cite{kar:88the}

Instead, it is preferable to examine the information contained in the
state vector by means of the study of an
observable computed from it. Methods in quantum chemical
topology (QCT), for instance, the Quantum Theory of Atoms in Molecules
(QTAIM)~\cite{libroBader} 
and the Interacting Quantum Atoms (IQA)~\cite{iqa2005,iqa2006}
energy partition are based on the exploitation and analysis of reduced
density matrices and which
have the attractive features of

\begin{itemize}
\item small basis set dependency (in a similar way to any other 3D
partition),
\item having orbital invariance,
\item providing the
division of molecular properties (particularly the electronic energy)
in physically sound components and
\item %IQA has the favorable property that it is
independence of the atomic virial theorem (only for IQA) which confers
applicability in every point of the configuration space of a given
electronic system~\cite{iqa2005,iqa2006}.
\end{itemize}

%These elements have enabled QTAIM and IQA to
%address many different chemical processes and systems on the same
%footing~\cite{safar,slaw,poulsen,abram,louis,andres,andres14,moa,ferr,tog,
%brovarets,cheg,mayer:01tca,salvador:01jcp,vyboishchikov:05jcp,tognetti:14pccp,salvador:04jcp}.
These conditions have enabled QTAIM to address many different chemical
processes and systems on the same
footing~\cite{safar,slaw,poulsen,abram,louis,andres,andres14,moa,ferr,tog,
brovarets,cheg,mayer:01tca,salvador:01jcp,vyboishchikov:05jcp,tognetti:14pccp,salvador:04jcp}.
In similar fashion,
 the IQA approach %an important
%method in QCT which 
has recently been applied to the study of
transition metal-ligand interactions~\cite{cukrowski2014,tiana2011,tiana2010}, 
bonding between electronegative atoms~\cite{tognetti2013}, 
the transferability of different species inside oligopeptides~\cite{maxwell}, 
the formation of water clusters~\cite{cukrowski2015,Guevara-Vela2013} 
and the conformational arrangement of carboxylic acids~\cite{ferro2015}.
%, among others.

The IQA energy partition has been implemented along with spin-independent
density matrices computed from Hartree-Fock (HF),~\cite{mayer:01tca,salvador:01jcp}
Complete Active Space Self Consistent Field (CASSCF)~\cite{iqa2005,iqa2006},
density functional theory (DFT),~\cite{vyboishchikov:05jcp}
and Full Configuration Interaction
wavefunctions~\cite{iqa2005,iqa2006}. Recently, dynamical correlation (DC) was included in
the IQA energy partition by means of closed shell (\textit{i}) HF/CC
transition density matrices~\cite{iqaCCDensTrans}
and (\textit{ii}) the coupled cluster singles and doubles (CCSD) lagrangian\cite{ccLag}. 
The last-mentioned developments make the IQA method suitable for the
study of phenomena in physical chemistry wherein DC is important, for
instance, in
non-covalent interactions and chemical
bonding~\cite{hexamerosAgua}.

%There are mostly two causes of electron correlation: the Pauli principle and 

Correlation in chemistry is mostly due to the Pauli antisymmetry principle and
the electron coulombic repulsion. Both mechanisms usually lead to larger interelectronic 
distances (with some exceptions~\cite{pearson:09mp}) but they affect 
electronic pairs differently. The Pauli principle is
imposed by forcing antisymmetry in the wavefunction. As a result,
electrons of like spin components experience a reduced probability of being at
short interelectronic distances; such effect is known as Fermi correlation.
On the other hand, the Coulomb repulsion among electrons influences
any pair of these particles regardless of their spin projection. The correlation effects
upon unlike-spin electron pairs are denominated as Coulomb correlation.

Single-determinant wavefunctions only consider Fermi
correlation, whereas Coulomb correlation is mainly DC and, therefore, it can be 
introduced by means of post-HF methods such as CC. 
However, a chief deficiency of coupled-cluster method is the difficulties it
creates for the calculation of molecular properties because the Hellmann-Feynman
theorem is not satisfied.~\cite{libroHelgaker} Namely, the definition of first-
and second-order matrices is not unique and, to our knowledge, all the available
expressions suffer from the $N$-representability problem.~\cite{coleman:00book}
Hence, the construction of appropriate
CC density matrices including correlation effects with minimal violation of the
$N$-representability conditions is important in order to obtain accurate CC
properties. 
Detailed analysis of the electron-correlation effects introduced 
by (approximate) CC densities is needed in order to identify the limitations of
the existing approximations and provide guidance for the construction of new CC
density matrix approximations. In this regard, the IQA energy partition
allows for a thorough analysis of the DC effects introduced by CC
aproximated density functions.

%Notwithstanding the
%accounting of DC by coupled cluster theory,
%CC methods do not possess a well-defined second-order matrix
%and, 
%therefore, it is unclear to which extent they give a good account of Fermi and
%Coulomb correlation effects.

%The consideration of coupled cluster spin-dependent matrix
%functions would enable the 
%study of Fermi and Coulomb correlation effects separately in electronic
%systems as put forward (or exploited) in this paper. \color{black}

Besides providing insights into the usefulness of CC matrices,
this work is aimed to further increase the
applicability of the IQA method (and consequently the arsenal of QCT tools) 
by considering its implementation with
the spin-de\-pend\-ent first-order reduced density matrix
and the pair density.
%$\varrho_1^{\sigma \sigma}(\mathbf{r}_1,\mathbf{r}_1^{\, \prime})$ and
%the pair density $\varrho_2^{\sigma \tau}(\mathbf{r}_1,
%\mathbf{r}_2)$ wherein $\sigma$ and $\tau$ denote either $\alpha$ or
%$\beta$ spin projection components. 
We believe that the use of these
spin-density matrices could be useful in quantum chemical topology
 and in general quantum
chemistry to investigate the effect of Fermi and Coulomb
correlation %(FCO and CCO) 
in different systems and processes, while they shed some light into
the electron correlation effects introduced by these approximate CC
density matrices.\newline
%In other words, we expect that the approach presented
%in this paper will prove useful in getting a deeper understanding of
%spin-dependent electron correlation in QCT and physical chemistry.

The rest of the article is organised as follows. We first describe
briefly the IQA energy partition. Then, we introduce the spin
contributions of the CC density matrices and 
the electron delocalisation indices, and
afterwards we give the computational details of the calculations
performed in this work. Finally, we discuss some illustrative
examples of the approach presented herein and present some
concluding remarks.
%effects

%in the study of ... and open shell
%systems.

\section{Interacting quantum atoms energy partition}
\label{IQA}
Different partitions of the three-dimensional space into (\textit{i}) disjoint
basins such as that provided by the quantum theory of atoms in
molecules or (\textit{ii}) interpenetrating densities as those suggested by
Becke~\cite{becke:88jcp,salvador:13jcp} and Hirshfeld~\cite{hirshfeld:77tca}
permit to divide the Born-Oppenheimer electronic
energy in monoatomic and diatomic terms,
\begin{align}
E &= \sum_{\mathrm{A}} E_{\mathrm{net}}^{\mathrm{A}} +
\frac{1}{2}\sum_{\mathrm{A \neq B}} 
E_{\mathrm{int}}^{\mathrm{A\cdots B}} \nonumber \\
 & = \sum_{\mathrm{A}} \left( T^{\mathrm{A}} + V_{\mathrm{ne}}^{\mathrm{AA}} + 
 V_{\mathrm{ee}}^{\mathrm{AA}} \right) + \frac{1}{2}\sum_{\mathrm{A \neq B}} 
\left( V_{\mathrm{nn}}^{\mathrm{AB}} + V_{\mathrm{ne}}^{\mathrm{AB}} +
V_{\mathrm{ne}}^{\mathrm{BA}} + V_{\mathrm{ee}}^{\mathrm{AB}} \right). \label{particion} 
\end{align}
\noindent $T^{\mathrm{X}}$ in equation (\ref{particion}) represents the kinetic energy of atom X,
while by letting $\gamma$ and $\delta$ to denote either electrons (e) or
nuclei (n), then $V_{\gamma \delta}^{\mathrm{XY}}$ indicates the
contribution to the potential
energy due to the interaction of $\gamma$ in atom X with $\delta$ in atom
Y. The expressions of $T^{\mathrm{X}}$ and $V_{\gamma
\delta}^{\mathrm{XY}}$  in terms of the reduced first order density
matrix $\varrho_1(\mathbf{r}_1; \mathbf{r}_1^{ \, \prime})$,
and the pair density $\varrho_2(\mathbf{r}_1, \mathbf{r}_2)$ 
are described in detail in Reference \mbox{[\!\!\citenum{iqa2005}]}. In order to
discuss the Fermi and Coulomb correlation into the
IQA partition energy, we have considered the non-vanishing spin
components of $\varrho_1(\mathbf{r}_1; \mathbf{r}_1^{ \, \prime})$ 
and $\varrho_2(\mathbf{r}_1, \mathbf{r}_2)$ for a state with a
definite value of $M_S$\cite{mcweenyMizuno1961} , \ie
\begin{align}
\varrho_1(\mathbf{r}_1; \mathbf{r}_1^{ \, \prime}) &= \varrho_1^{
\alpha \alpha }(\mathbf{r}_1; \mathbf{r}_1^{ \, \prime})
+ \varrho_1^{ \beta \beta }(\mathbf{r}_1; \mathbf{r}_1^{ \, \prime}), \label{rho1Spin}  \\
\varrho_2(\mathbf{r}_1, \mathbf{r}_2) &= \varrho_2^{ \alpha \alpha }(\mathbf{r}_1, \mathbf{r}_2) + 
\varrho_2^{ \alpha \beta }(\mathbf{r}_1, \mathbf{r}_2) +
\varrho_2^{ \beta \alpha }(\mathbf{r}_1, \mathbf{r}_2) +
\varrho_2^{ \beta \beta }(\mathbf{r}_1, \mathbf{r}_2). \label{rho2Spin}
\end{align}
\noindent The spin-configurations in the RHS of equations
(\ref{rho1Spin}) and (\ref{rho2Spin}) are those that contribute to
the calculation of
expectation values of the spin-independent 
electronic Hamiltonian.~\cite{McWeeny1992} 

The IQA interaction energies can also be further
divided by considering the Coulombic and exchange-correlation
components of the pair density~\cite{iqa2005}
\begin{align}
\varrho_2(\mathbf{r}_1,\mathbf{r}_2) &=
\varrho_2^{\mathrm{J}}(\mathbf{r}_1,\mathbf{r}_2) +
\varrho_2^{\mathrm{xc}}(\mathbf{r}_1,\mathbf{r}_2) \nonumber \\
& =
\varrho(\mathbf{r}_1) \varrho(\mathbf{r}_2) +
\varrho_2^{\mathrm{xc}}(\mathbf{r}_1,\mathbf{r}_2), \label{rho2_XC_J}
\end{align}
\noindent into a classical, \ie electrostatic
\begin{align}
V_{\mathrm{cl}}^{\mathrm{AB}} &= 
V_{\mathrm{nn}}^{\mathrm{AB}} + V_{\mathrm{ne}}^{\mathrm{AB}} +
V_{\mathrm{ne}}^{\mathrm{BA}} + V_{\mathrm{J}}^{\mathrm{AB}},
\end{align}
\noindent  and a quantum-mechanical (exchange-correlation) contribution 
$V_{\mathrm{xc}}^{\mathrm{AB}}$, in a way that~\cite{iqa2005}
\begin{align}
E_{\mathrm{int}}^{\mathrm{AB}} &=
V_{\mathrm{cl}}^{\mathrm{AB}}  +
V_{\mathrm{xc}}^{\mathrm{AB}}.
\end{align}
As stated before, we will be concerned in this article with the spin-components of the pair density
\begin{align}
\varrho_2^{\sigma \tau}(\mathbf{r}_1,\mathbf{r}_2) &=
\varrho_2^{\sigma \tau, \, \mathrm{J}}(\mathbf{r}_1,\mathbf{r}_2) +
\varrho_2^{ \sigma \tau, \, \mathrm{xc}}(\mathbf{r}_1,\mathbf{r}_2) \nonumber \\
& =
\varrho^{\sigma}(\mathbf{r}_1) \varrho^{\tau}(\mathbf{r}_2) +
\varrho_2^{ \sigma \tau, \ \mathrm{xc}}(\mathbf{r}_1,\mathbf{r}_2),
\label{rho2_XC_J_Espin}
\end{align}
\noindent in which $\sigma$ and $\tau$ each indicates an $\alpha$ or
$\beta$ spin projection. 
The spin-dependent density
matrices in formulae (\ref{rho1Spin})--(\ref{rho2Spin}) 
will be exploited to assess separately the Fermi and Coulomb correlation
effects on the net and interatomic components of the IQA partition as
discussed in the next section.

\section{Spin-dependent one- and two-electron matrices}
\label{spin2RDM}

%\color{red} Creo que lo que hay que indicar es que se van a considerar las spin-dependent matrices HF
%y CC. \color{black}

%\color{red} No estoy seguro que se usen las HF spin-dependent\ldots
%In the remaining part of the paper 
We will consider only closed-shell systems and
thus the expressions used in this section to take into account DC 
are only valid in this context.
The HF spin-dependent density matrices read
%\color{red} Expresiones de Helgaker \color{black}

\hspace*{-1.0cm}\begin{minipage}[t]{1.15\textwidth}
\begin{align}
\varrho_1^{\sigma \sigma, \, \HF}(\mathbf{r}_1; \ \mathbf{r}_1^{\, \prime}) &= \sum_{p}
k_{p\sigma} \varphi_{p}^{\, \star}(\mathbf{r}_1^{\, \prime})
\varphi_{p}(\mathbf{r}_1), \label{Dpq_SpinHF}  \\
\varrho_2^{\sigma \tau, \, \HF}(\mathbf{r}_1, \ \mathbf{r}_2) &=
\sum_{pq} k_{p\sigma} k_{q\tau} \left[
|\varphi_p(\mathbf{r}_1)|^2
|\varphi_q(\mathbf{r}_2)|^2 - \delta_{\sigma \tau}  \varphi_p^{\, \star}(\mathbf{r}_1) \varphi_p(\mathbf{r}_2)
\varphi_q^{\, \star}(\mathbf{r}_2) \varphi_q(\mathbf{r}_1) \right],
\label{dpqrs_SpinHF}
\end{align}
\end{minipage}

\vspace*{0.5cm}

\noindent in which $\sigma$ and $\tau$ have the same meaning that in
equation (\ref{rho2_XC_J_Espin}), $\{\varphi_p(\mathbf{r})\}$ is the
set of spatial molecular orbitals used to construct the Fock space of
the system under consideration, $k_{p\sigma}$ represents the
occupation number of spin orbital $\varphi_p(\mathbf{r})\sigma(s)$ in
$| \HF \rangle$ and $\delta_{\sigma \tau}$ denotes the Kronecker delta. \color{black} 
Equation (\ref{dpqrs_SpinHF}) can be rewritten entirely in terms of 
expression (\ref{Dpq_SpinHF}), \ie,

\hspace*{-1.0cm}\begin{minipage}[t]{1.1\textwidth}
\begin{equation}\label{HFL}
\varrho_2^{\sigma \tau, \, \HF}(\mathbf{r}_1, \ \mathbf{r}_2) =
\varrho^{\sigma, \, \HF}(\mathbf{r}_1)\varrho^{\tau, \, \HF}(\mathbf{r}_2)-\delta_{\sigma \tau}
\varrho_1^{\sigma \sigma, \, \HF}(\mathbf{r}_1; \ \mathbf{r}_2)
\varrho_1^{\sigma \sigma, \, \HF}(\mathbf{r}_2; \ \mathbf{r}_1).
\end{equation}
\end{minipage}

\vspace*{0.5cm}

\noindent The last equation shows that the HF method
does not include unlike-spin contributions in the pair density beyond the indepedent-pair
distribution, $\varrho^{\sigma, \, \HF}\varrho^{\tau, \, \HF}$,
and, therefore, does
not contain any Coulomb correlation. The same is true for
 HF-like approximations to the
pair density, which for a given correlated method use the expression (\ref{HFL}) to estimate
$\varrho_2(\mathbf{r}_1,\mathbf{r}_2)$ 
but replace $\varrho^{\sigma, \, \HF}$
and $\varrho_1^{\sigma \sigma, \, \HF}$ with the pertinent correlated counterparts.

Along with the scalar fields
$\varrho_1^{\sigma
\sigma, \, \HF}$ and $\varrho_2^{\sigma \tau, \, \HF}$ defined in
equations (\ref{Dpq_SpinHF}) and (\ref{dpqrs_SpinHF}), we will take
into account the corresponding functions based in
HF/CC transition density matrices. As established in reference 
[\!\!\citenum{iqaCCDensTrans}], the scalar fields
\begin{align}
\rho_1^{\HF/\CC}(\mathbf{r}_1; \ \mathbf{r}_1^{\, \prime}) & =
\sum_{pq} D_{pq}^{\HF/\CC} \varphi^{\, \star}_{p} (\mathbf{r}_{1}^{\,
\prime}) \varphi_{q}(\mathbf{r}_{1}) \nonumber \\
& = \sum_{pq} \langle \HF | \E_{pq} | \CC \rangle 
\varphi^{\, \star}_{p} (\mathbf{r}_{1}^{\, \prime}) \varphi_{q}
(\mathbf{r}_{1}), \label{densTrans1} \\
\phantom{a} & \phantom{b} \nonumber \\
\rho_2^{\HF/\CC} (\mathbf{r}_1, \ \mathbf{r}_2) & =
\sum_{pqrs} d_{pqrs}^{\HF/\CC} 
\varphi^{\, \star}_{p}(\mathbf{r}_{1}) \varphi_{q} (\mathbf{r}_{1}) 
\varphi^{\, \star}_{r}(\mathbf{r}_{2}) \varphi_{s} (\mathbf{r}_{2})
  \nonumber \\
& = \sum_{pqrs} \langle \HF | \e_{pqrs} | \CC \rangle 
\varphi^{\, \star}_{p}(\mathbf{r}_{1}) \varphi_{q} (\mathbf{r}_{1}) 
\varphi^{\, \star}_{r}(\mathbf{r}_{2}) \varphi_{s} (\mathbf{r}_{2}),
\label{densTrans2}
\end{align}
%\noindent wherein $\displaystyle \E_{pq} = \ad_{p \alpha} \an_{q \alpha} + \ad_{p \beta} \an_{q \beta}$ and 
%$\displaystyle \e_{pqrs} = \E_{pq} \E_{rs} - \delta_{qr} \E_{ps}$, 
%\vspace*{-0.3cm}
\noindent can be used to include electron correlation in the
IQA energy partition of closed shell species. 
The quantities $D_{pq}^{\HF/\CC}$ and $d_{pqrs}^{\HF/\CC}$ in the RHS
of equations (\ref{densTrans1}) and (\ref{densTrans2}) are one- and
two-electron matrices used to obtain the first and second-order
density functions respectively, while 
\begin{align}
\E_{pq} & = \ad_{p \alpha} \an_{q \alpha} + \ad_{p
\beta} \an_{q \beta}, \ \mbox{and} \label{Epq} \\
\e_{pqrs} & = \E_{pq} \E_{rs} - \delta_{qr} \E_{ps}. \label{epqrs} 
\end{align}
The spin components of
the density functions (\ref{densTrans1}) and (\ref{densTrans2}) are 
%\color{red} Expresiones de la p\'agina 5 del escrito a mano \color{black}
\begin{align}
\rho_1^{\sigma \sigma, \, \HF/\CC}(\mathbf{r}_1; \ \mathbf{r}_1^{\,
\prime}) & =  \sum_{pq} D_{p \sigma q \sigma}^{\HF/\CC} 
\varphi^{\, \star}_{p} (\mathbf{r}_{1}^{\, \prime})
\varphi_{q}(\mathbf{r}_{1}) \nonumber \\
 &= \sum_{pq} \langle \HF | \ad_{p \sigma} \an_{q \sigma} | \CC \rangle 
\varphi^{\, \star}_{p} (\mathbf{r}_{1}^{\, \prime}) \varphi_{q}
(\mathbf{r}_{1}), \label{densTransEspin1} \\
\phantom{a} & \phantom{b} \nonumber \\
\vspace*{-1.0cm}
\rho_2^{\sigma \tau, \, \HF/\CC} (\mathbf{r}_1, \ \mathbf{r}_2) & =
\sum_{pqrs} d_{p \sigma q \sigma r \tau s \tau}^{\HF/\CC}
 \varphi^{\, \star}_{p}(\mathbf{r}_{1}) \varphi_{q} (\mathbf{r}_{1}) 
\varphi^{\, \star}_{r}(\mathbf{r}_{2}) \varphi_{s} (\mathbf{r}_{2}),
\nonumber \\
& = \sum_{pqrs} \langle \HF | \left( \ad_{p \sigma} \an_{q \sigma} \ad_{r
\tau} \an_{s \tau} - \delta_{\sigma \tau} \delta_{qr}\ad_{p \sigma}
\an_{s \sigma} \right) | \CC \rangle \nonumber \\  
 & \phantom{= \sum_{pqrs}} \ \  \times \varphi^{\, \star}_{p}(\mathbf{r}_{1}) \varphi_{q} (\mathbf{r}_{1}) 
\varphi^{\, \star}_{r}(\mathbf{r}_{2}) \varphi_{s} (\mathbf{r}_{2}).
\label{densTransEspin2}
\end{align}
The matrix elements within equations (\ref{densTransEspin1}) and
(\ref{densTransEspin2}) can be computed according to equations:
%\color{red} Las ecuaciones que est\'an en el sobre \color{black}

\hspace*{-1.0cm}\begin{minipage}[t]{1.25\textwidth}
\begin{align}
\langle \HF | \ad_{p \sigma} \an_{q \sigma} | \CC \rangle & = \left\{ 
\begin{array}{ll} 
\delta_{pq} & \mbox{if $p \in$ occ; $q \in$ occ} \\ 
t_{p}^{q} & \mbox{if $p \in$ occ; $q \in$ virt} \\ 
0 & \mbox{in any other case.} 
\end{array} \right.  \label{apaq} \\
\phantom{a} & \phantom{b} \nonumber \\
\langle \HF | \ad_{p \sigma} \an_{q \sigma} \ad_{r \tau} \an_{s \tau}
| \CC \rangle & = \left\{ 
\begin{array}{ll}
\delta_{pq}\delta_{rs} & \mbox{if $p, \ q, \ r, \ s \in$ occ} \\
\delta_{pq}t_r^s & \mbox{if $p, \ q, \ r \in$ occ; $s \in$ vir} \\
\delta_{rs}t_p^q -  \delta_{\sigma \tau} \delta_{ps} t_r^q & \mbox{if $p, \ r, \ s \in$ occ; $q \in$ vir} \\
t_p^q t_r^s + t_{pr}^{qs} - \delta_{\sigma \tau} (t_p^s t_r^q +
t_{pr}^{qs})  & \mbox{if $p, \ r \in$ occ; $q, \ s \in$ vir} \\
\delta_{\sigma \tau} \delta_{qr}\delta_{ps} & \mbox{if $p, \ s \in$ occ; $q, \ r  \in$ vir} \\
\delta_{\sigma \tau} \delta_{qr} t_p^s & \mbox{if $p \in$ occ; $q, \ r, \ s  \in$ vir} \\
0 & \mbox{in any other case.}
\end{array} \right. \label{apaqaras}
\end{align}
\end{minipage}

%...WAIT A MINUTE... CREO QUE ESTA CUMPLE MÁS
%SIMETRÍAS, HAY QUE REVISAR LAS DEL CAPÍTULO 1 DE HELGAKER
\vspace*{0.5cm}

Since expressions (\ref{densTrans1})--(\ref{apaqaras}) refer to
closed-shell coupled-cluster theory, these equations are symmetric in the
$\sigma$ and $\tau$ spin projections, \ie,
\begin{align} 
\langle \HF | \ad_{p \alpha} \an_{q \alpha} | \CC \rangle &= 
\langle \HF | \ad_{p \beta} \an_{q \beta} | \CC \rangle,
\label{simetriaEspin1} \\
\langle \HF | \ad_{p \sigma} \an_{q \sigma} \ad_{r \tau} \an_{s \tau}
| \CC \rangle & = 
\langle \HF | \ad_{p \tau} \an_{q \tau} \ad_{r \sigma} \an_{s \sigma}
| \CC \rangle.
\label{simetriaEspin2}
\end{align}
By taking into consideration the symmetry relations~\cite{libroHelgaker} 
\begin{align}
D_{p\sigma q\sigma} &=  D_{q\sigma p\sigma},  \label{simLaDeUno} \\
d_{p\sigma q \sigma r \tau s \tau}& = 
d_{r\tau s \tau p \sigma q \sigma}= 
d_{q\sigma p \sigma s \tau r \tau} = 
d_{s\tau r \tau q \sigma p \sigma}, 
\end{align}
\noindent wherein it is assumed that the molecular orbitals used to
construct the $| \HF \rangle$ and $| \CC \rangle$ approximate
wavefunctions are real, we obtain the spin-dependent one- and
two-electron matrices

%\vspace*{-0.3cm}

\hspace*{-1.0cm}\begin{minipage}[t]{1.15\textwidth}
\begin{align}
D_{i\sigma j\sigma}^{\HF/\CC} &= \delta_{ij}, \label{dij} \\
D_{i \sigma a \sigma}^{\HF/\CC} & = D_{a \sigma i \sigma}^{\HF/\CC} 
 = \frac{t_i^a}{2}, \label{dia} \\
d_{i \sigma j \sigma k \tau  l \tau}^{\HF/\CC} &= 
d_{k \tau l \tau i \sigma  j \sigma}^{\HF/\CC} = 
d_{j \sigma i \sigma l \tau  k \tau}^{\HF/\CC} = 
d_{l \tau k \tau j \sigma  i \sigma}^{\HF/\CC} = 
\delta_{ij} \delta_{kl} - \delta_{\sigma \tau} \delta_{jk}\delta_{il}, \label{dijkl} \\
d_{i \sigma j \sigma k \tau  a \tau}^{\HF/\CC} &= 
d_{k \tau a \tau i \sigma  j \sigma}^{\HF/\CC} = 
d_{j \sigma i \sigma a \tau  k \tau}^{\HF/\CC} = 
d_{a \tau k \tau j \sigma  i \sigma}^{\HF/\CC} = 
\frac{1}{2}\left( \delta_{ij}t_k^a -
\delta_{\sigma \tau} \delta_{k j}t_i^a \right), \label{dijka} \\
d_{i \sigma a \sigma j \tau  b \tau}^{\HF/\CC} &= 
d_{j \tau b \tau i \sigma  a \sigma}^{\HF/\CC} = 
d_{a \sigma i \sigma b \tau  j \tau}^{\HF/\CC} = 
d_{b \tau j \tau a \sigma  i \sigma}^{\HF/\CC} = 
\frac{1}{2}\left( t_i^a t_j^b + t_{ij}^{ab} -\delta_{\sigma \tau}
\left( t_i^b t_j^a +  t_{ij}^{ba} \right) \right), \label{diajb} 
\end{align}
\end{minipage}

\vspace*{0.5cm}

\noindent in which $i, \ j, \ k \ldots$ ($a, \ b, \ c \ldots$)
represent HF occupied (virtual) orbitals in ac\-cord\-ance with common
use. Although the antepenultimate and penultimate rows of equation
(\ref{apaqaras}) suggest that we have non-vanishing blocks
$d_{i \sigma a \sigma b \tau j \tau}^{\HF/\CC}$
and $d_{i \sigma a \sigma b \tau c \tau}^{\HF/\CC}$, that is indeed, not the case
%equations (\ref{apaq}) and (\ref{apaqaras})
\begin{align}
d_{i \sigma a \sigma b \tau j \tau}^{\HF/\CC} &= \langle \HF | \ad_{i \sigma}
\an_{a \sigma} \ad_{b \tau} \an_{j \tau} | \CC \rangle -
\delta_{\sigma \tau} \delta_{ab} \langle \HF | \ad_{i \sigma} \an_{j
\tau} | \CC \rangle \nonumber \\
 &= \delta_{\sigma \tau} \delta_{ab} \delta_{ij} -
 \delta_{\sigma \tau} \delta_{ab} \delta_{ij} = 0, \nonumber \\
%\phantom{a} & \phantom{b} \nonumber \\
d_{i \sigma a \sigma b \tau c \tau}^{\HF/\CC} &= \langle \HF | \ad_{i \sigma}
\an_{a \sigma} \ad_{b \tau} \an_{c \tau} | \CC \rangle -
\delta_{\sigma \tau} \delta_{ab} \langle \HF | \ad_{i \sigma} \an_{c
\tau} | \CC \rangle \nonumber \\
 &= \delta_{\sigma \tau} \delta_{ab} t_i^c -
 \delta_{\sigma \tau} \delta_{ab} t_i^c = 0. \nonumber 
\end{align}
%, we
%obtain,
%
%\color{red} ecuaciones simetrizadas \color{black}
%\vspace*{0.3cm}
In Mcweeny's normalization~\cite{McWeeny1992}, the spin-dependent pair density 
reduces to the spin-dependent density upon integration of one coordinate
%
%\vspace*{-0.2cm}
%
\begin{align}
\sum_\tau \int \rho_2^{\sigma \tau} (\mathbf{r}_1, \mathbf{r}_2)
\mathrm{d} \mathbf{r}_2 & = (N - 1)
\rho^{\sigma}(\mathbf{r}_1), \label{rho1Rho2Int}
\end{align}
%
%\vspace*{0.2cm}
%
\noindent where $N$ is the number of electrons of the system. Equation
(\ref{rho1Rho2Int}) implies that
\begin{align}
\sum_{r \tau} d_{p \sigma q \sigma r \tau r \tau} & = (N - 1)
D_{p\sigma q\sigma}. \label{rho1Rho2Sum}
\end{align}
\noindent It is not complicated to verify that the one- and
two-electron matrices $D_{p \sigma q \sigma}^{\HF/\CC}$ and 
$d_{p \sigma q \sigma r \tau s \tau}^{\HF/\CC}$ in equations
(\ref{dia})--(\ref{diajb}) fulfil condition
(\ref{rho1Rho2Sum}).

Formulae %(LAS DE LOS SPIN-DEPENDENT ONE AND TWO ELECTRON MATRICES) 
(\ref{Dpq_SpinHF}) and (\ref{dpqrs_SpinHF}) along with the substitution
of expressions (\ref{dij})--(\ref{diajb}) in equations 
(\ref{densTransEspin1}) and (\ref{densTransEspin2}) 
are used in this work to investigate 
%the importance of FC and CC 
Fermi and Coulomb correlation effects
in the IQA energy partition as illustrated
in Section \ref{results}. IQA analyses are often accompanied by an
examination of delocalisation indices (DI) which are briefly reviewed
in the next section.
 
%\color{red} 
%COMPLETAR LOS HUECOS Y LLENAR CON ALGO MÁS DE LAS COSAS QUE DEBAN SATISFACER ESTAS MADRES.

\section{Delocalisation Indices}
\label{DIs}

Population analysis comprises a set of techniques that assign a number of electrons, 
the atomic population, to each atom in an electronic system, affording a means to
distribute the $N$ electrons in a molecule or molecular cluster among
their constituent parts~\cite{mulliken:55jcp}.
The atomic population in the QTAIM is defined solely from the electron 
density~\cite{libroBader},

\begin{align}
N_{\text{A}} & =\int_{\text{A}} \rho(\mathbf{r}_1) \mathrm{d}\mathbf{r}_1,
\label{na}
\end{align}

\noindent where A is the corresponding QTAIM atom, and 

\begin{align}\label{sumrule}
N & =\sum_{\text{A}} N_{\text{A}},
\end{align}

\noindent in which $N$ is the number of electrons in the system.
The variance and 
covariance of atomic populations lead to the definition of localisation (LI)
and delocalisation indices (DI) ~\cite{bader:74cpl,bader:75jacs,fradera:99jpca}

\begin{align}
\lambda^{\text{A}} & = N_{\text{A}}-\sigma^2\left[N_{\text{A}}\right], \label{lambda} \\
\delta^{\text{AB}} & =2\left(N_{\text{A}}N_{\text{B}}-\left<N_{\text{A}}N_{\text{B}}\right>\right),
\label{delta}
\end{align}

\noindent wherein

\begin{align}
\sigma^2\left[N_{\text{A}}\right] & =\left<N_{\text{A}}^2\right>-N_{\text{A}}^2, \\
\left<N_{\text{A}}N_{\text{B}}\right> & =\int_{\text{A}}\int_{\text{B}}\rho_2(\mathbf{r}_1,\mathbf{r}_2)
\mathrm{d}\mathbf{r}_1\mathrm{d}\mathbf{r}_2 +N_{\text{A}}\delta_{\text{AB}}.
%\left<N_{\text{A}}^2\right>=\int_{\text{A}}\int_{\text{A}}\rho_2(\mathbf{r}_1,\mathbf{r}_2)
%\mathrm{d}\mathbf{r}_1\mathrm{d}\mathbf{r}_2+N_{\text{A}}
\end{align}

\noindent where $\delta_{AB}$ is a Kronecker delta.
One can easily prove that the following property %is fulfilled:

\begin{align}
N_{\text{A}} & =\lambda^{\text{A}}+\frac{1}{2}\sum_{\text{B}\neq \text{A}} \delta^{\text{AB}},
\end{align}
is attained.
Following this scheme one can decompose the number of electrons in a 
system into atomic regions (equation (\ref{na})).
In turn, it is possible to divide atomic populations into 
electrons localised in atom A (expression (\ref{lambda})) 
or delocalised between atom A and the other atoms in the 
molecule (formula (\ref{delta})), using not only QTAIM 
but any other atomic partition~\cite{matito:05jpca}.
In principle, the latter decomposition depends on the pair density,
and therefore a considerable computational effort is required to perform it.
Hence, several approximations
to the DI have been suggested~\cite{wang:03jcc,matito:07fd,dmftiqa,feixas:10jctcelf}.
Here we study the two most popular ones, based on 
M\"uller's approximation to the pair density~\cite{muller:84pl},  
which gives rise to Fulton's definition of the
electron sharing index~\cite{fulton:93jpc}, $\delta^{\text{AB}}_{F}$ 
and the Hartree-Fock-like approximation~\cite{lowdin:55pr} of the pair
density (equation \ref{HFL}) that
leads to the DI proposal of {\'A}ngy{\'a}n's and coworkers~\cite{angyan:94jpc}, 
$\delta^{\text{AB}}_{A}$.
The latter cannot contain Coulomb correlation effects as pointed out
in the text below equation (\ref{HFL}), whereas the former
has been shown to provide a good account of both Fermi and Coulomb correlation
effects in configuration interaction singles and doubles (CISD),~\cite{matito:07fd,feixas:10jctcelf}
and ground-state~\cite{wang:03jcc} and excited states~\cite{feixas:11pccp} CASSCF wavefunctions.
In this work we will compare these approximations with the Fermi and Coulomb
parts of the DI, \ie

\begin{align}\label{DIdecomp}
\delta^{\text{AB}}= \sum_{\sigma} \delta^{\text{AB},\sigma\sigma} +
\sum_{\sigma \neq \tau} \delta^{\text{AB},\sigma \tau}.
\end{align}

%\noindent where $\sigma \neq \tau$.

\section{Computational details}
\label{cdetails}

%ºThe assessment of the Fermi and Coulomb correlation in the IQA energy
%ºpartition is done by considering the systems \ce{HC#CH\bond}, \ce{CN^-},

The use of spin-dependent matrices in the IQA energy partition
proposed in this work is illustrated by considering 
\ce{HC#CH}, \ce{BeH2}, \ce{BH}, \ce{CN-}, \ce{HF}, \ce{NO+}, \ce{LiH}, 
\ce{LiF} and \ce{H2O\bond{...}H2O} and 
which comprise non-polar 
covalent, polar covalent, ionic and hydrogen bonded systems.
%In addition, we carried the electronic partition
%addressed in this work for the free radicals \ce{CH3.}, \ce{CH3CH2.},
%\ce{(CH3)2CH.}, \ce{(CH3)3C.}. The open shell species were considered
%only with HF spin dependent matrices (equations (\ref{Dpq_SpinHF}) and
%(\ref{dpqrs_SpinHF})) whereas the
%electronic energies of the closed shell species were also 
%addressed with spin-dependent HF/CC transition densities (expressions
%(\ref{dia})--(\ref{diajb})).
This will allow us to assess the effects of Fermi and Coulomb
correlation in the IQA energy partition in different chemical
situations. The geometries of all systems were optimised with the
CCSD/cc-pVTZ approximation (apart from the water dimer for which we
carry out a CCSD/aug-cc-pVTZ geometry optimisation) as implemented in
{\sc{Gaussian-09}}~\cite{g09B}. Later, we carried out single point calculations to
procure the coupled cluster amplitudes necessary to compute the HF/CC
transition densities (formulae \ref{dia}--\ref{diajb}) with the
quantum chemistry package {\sc{Molpro}}~\cite{molpro,ccsdMolpro1,ccsdMolpro2}.

Once computed the matrices
$\mathbf{D}^{\HF/\CC}_{\sigma \sigma}$, $\mathbf{D}^{\HF/\CC}_{\sigma
\tau}$,
$\mathbf{d}^{\HF/\CC}_{\sigma \sigma}$, $\mathbf{d}^{\HF/\CC}_{\sigma \tau}$ 
put forward in this work, 
we used the software {\sc{Imolint}}~\cite{imolint} to determine
the total molecular
electronic energy in terms of these one- and two-electron matrices.
The same program was used to calculate the spin-dependent electron-electron 
repulsion, exchange and
correlation contributions of the whole electronic systems 
prior to carry out the IQA electronic energy partition.
The IQA analysis was performed
with the code {\sc{Promolden}}~\cite{promolden} using the QTAIM zero-flux surface to
divide the three-dimensional space of the system. 
We considered
%(\textit{i}) $\beta$ spheres with radii equal to 90/60 \% \color{olive}{\'Ise:
%{?`}siempre en cu\'anto qued\'o el asunto?} \color{black} of the distance to the
%closest bond critical point 
($i$) $\beta$-spheres with radii that were partially optimised, starting from
our standard prescription that 
equates them to 90\% the distance from a nucleus to its closest bond
critical point, 
along with (\textit{ii}) a considerable
large number of radial and angular integration grids
to get a suitable numerical precision for
the IQA integrations. 
More specifically, 
numerical integrations were performed using large 5810
points Lebedev angular grids and $l = 10$ spherical harmonics
expansions. Radial parameters were precision oriented. With this we
mean that they were selected so as to warrant meaningful precision in
the energetic quantities here presented. Since $\beta$-spheres were used in
all the cases, two (inner/outer) radial grids had to be chosen. 
Some difficult systems required 
900/800 points while most were found to be reasonably integrated with 
400/400 or even 200/200 grids.
In the last two cases, the inner $l$ expansion was cut at
$l = 6$. 
%The $\beta$-spheres radii were partially optimised, starting from our
%standard prescription that equates them to 90\% the distance from a
%nucleus to its closest bond critical point. %H$_2$S was shown to require very stringent
%computational conditions, with 2000/2000 radial grids. 

Finally, we used the {\sc{ESI-3D}} program~\cite{esi3d} to calculate the genuine,
the approximated and the decomposition of the DIs in its like and
unlike spin contributions using the atomic
overlap matrices provided by {\sc{Promolden}}.

\section{Results and discussion}
\label{results}

%\documentclass[a4paper,8pt]{article}
%\usepackage[utf8]{inputenc}
%
%\usepackage[version=4]{mhchem}
%
%%opening
%\title{}
%\author{}
%
%%\begin{document}
%
%\maketitle
%
%%
%\section{}

\begin{table}
\caption{Differences between the total electronic energies computed 
with (\textit{i}) {\sc{Imolint}} and (\textit{ii}) {\sc{Promolden}} as
compared with those obtained with the \emph{ab initio} package
{\sc{Molpro}}. The data are reported in milliHartree.}
\begin{center}
\begin{tabular}{l|r@{.}l|r@{.}l|r@{.}l|r@{.}l|r@{.}l}
\hline
 & \multicolumn{4}{|c|}{\sc{Imolint}} &
\multicolumn{4}{|c}{\sc{Promolden}} \\
\cline{2-9}
& \multicolumn{2}{c|}{HF} &
\multicolumn{2}{c|}{HF/CC} 
& \multicolumn{2}{c|}{HF} &
\multicolumn{2}{c}{HF/CC}  \\
\hline
 %\ce{LiF} &   $-$4&467E-03   &  0&000  &  0&000  &  0&000 \\
%\hline
\ce{BeH2} &  $-$1&$40 \times 10^{-4}$ &  $-$7&$10 \times 10^{-4}$ &  0&19  &  0&57 \\
\ce{BH}   &  $-$4&$20 \times 10^{-4}$ &  1&$26 \times 10^{-2}$ &  $-$2&$58 \times 10^{-2}$  &  $-$4&$54 \times 10^{-3}$ \\
\ce{CN-}  &  $-$3&$00 \times 10^{-5}$ &  $-$4&$20 \times 10^{-4}$ & $-$0&19  & $-$0&38 \\
\ce{HF}   &   6&$20 \times 10^{-4}$ &  3&$38 \times 10^{-3}$ &  $-$0&35  &  0&42 \\
\ce{LiF}  &   4&$47 \times 10^{-3}$ &  1&$07 \times 10^{-3}$ &  0&14  &  0&12 \\
\ce{NO+}  &   4&$60 \times 10^{-4}$ &  $-$1&$74 \times 10^{-2}$ &
0&26  &  $-$0&15 \\
\ce{LiH}  &   2&$71 \times 10^{-3}$ &  2&$32 \times 10^{-3}$ &  $-$4&$63 \times 10^{-2}$  &  $-$0&05 \\
\ce{H2O\bond{...}H2O}  & $-$1&$60 \times 10^{-4}$ &  3&$83 \times
10^{-3}$ & $-6$&11  &  $-3$&99 \\
\ce{HC#CH} & $-$1&$80 \times 10^{-4}$ &  3&$99 \times 10^{-3}$ &  0&82  &  1&02 \\
%cambiar los nombres de los radicales
\hline
%\ce{CH3.} &  2&$64 \times 10^{-4}$  & \multicolumn{2}{c|}{---}  &
%6&$94 \times 10^{-2}$  & \multicolumn{2}{c|}{---}  \\
%\ce{CH3CH2.} &  6&$27 \times 10^{-4}$  & \multicolumn{2}{c|}{---}  &
%$-$0&22  & \multicolumn{2}{c|}{---}  \\
%\ce{(CH3)2CH.} &  $-$4&$50 \times 10^{-5}$  & \multicolumn{2}{c|}{---}  &  $-$1&40  & \multicolumn{2}{c|}{---}  \\
%\ce{(CH3)3C.} &  $-$1&$71 \times 10^{-4}$  & \multicolumn{2}{c|}{---} &  $-$11&80  & \multicolumn{2}{c|}{---}  \\
%\hline
\end{tabular}
\end{center}
\label{tab:imolint}
%\end{document}
\end{table}

Table \ref{tab:imolint} shows the differences of the 
total energies computed with (\textit{i}) {\sc{Imolint}}
and {\sc{Promolden}} with either the HF and HF/CC spin-dependent
matrices and (\textit{ii})
the corresponding \emph{ab initio} results. We observed that the
discrepancies between {\sc{Molpro}} and {\sc{Imolint}} results are in the scale of
microHartrees while the order of magnitude of the integration errors
of {\sc{Promolden}} is below the range of milliHartrees. These
results show that the electronic energy can indeed be reproduced from
equations (\ref{Dpq_SpinHF})--(\ref{dpqrs_SpinHF}) and
(\ref{densTransEspin1}), (\ref{densTransEspin2}) in conjunction with (\ref{dij})--(\ref{diajb}), thereby indicating the suitability of these
spin-dependent one- and two-electron matrices to carry out the energy
partition of the systems addressed in this investigation. 

%We point %out that the integration error of \sc{Promolden} is
%usually much smaller than the quantities the components of the IQA
%energy partition. Although the 

%\input{tablaDeltasTotal}

\begin{table}

\caption{Changes due to the consideration of dynamic correlation by
means of HF/CC transition matrices in the electron-electron potential energy
($\Delta V_{\mathrm{ee}}$)
along with its same and different-spin contributions 
$\Delta V_{\mathrm{ee}}^{\sigma \sigma}$ and $\Delta V_{\mathrm{ee}}^{\sigma \tau}$
($\sigma \neq \tau$) 
for the molecules addressed in this work. 
The corresponding values for exchange
($\Delta V_{\mathrm{X}}$)
together with those of the spin-dependent correlation terms
$V_{\mathrm{corr}}^{\sigma \sigma}$ and 
$V_{\mathrm{corr}}^{\sigma \tau}$ are also reported.
Atomic units are used throughout.}

\begin{center}
%\hspace*{-2.0cm}
\begin{tabular}{l|r@{.}l|r@{.}l|r@{.}l|r@{.}l|r@{.}l|r@{.}l}
\hline
System & \multicolumn{2}{c|}{$ \Delta V_{\mathrm{ee}}$} & 
\multicolumn{2}{c|}{$\Delta V_{\mathrm{ee}}^{\sigma \sigma}$} &
\multicolumn{2}{c|}{$\Delta V_{\mathrm{ee}}^{\sigma \tau}$} & 
\multicolumn{2}{c|}{$\Delta V_{\mathrm{X}}$} & 
\multicolumn{2}{c|}{$V_{\mathrm{corr}}^{\sigma \sigma}$} &
\multicolumn{2}{c}{$V_{\mathrm{corr}}^{\sigma \tau}$} \\
\hline
\ce{BeH2} &  $-$0&079292  &  0&001350  &  $-$0&080642  &  $-$0&000122  &  $-$0&001578  &  $-$0&083693\\
\ce{BH} &  $-$0&076244  &  0&009250  &  $-$0&085494  &  $-$0&002357  &  $-$0&006541  &  $-$0&103642\\
\ce{CN-} &  $-$0&234918  &  $-$0&016746  &  $-$0&218172  &  0&003493  &  $-$0&072248  &  $-$0&269734\\
\ce{HF} &  $-$0&324188  &  $-$0&085156  &  $-$0&239032  &  0&000246  &  $-$0&042453  &  $-$0&196084\\
\ce{LiF} &  $-$0&418864  &  $-$0&119426  &  $-$0&299438  &  0&003659  &  $-$0&033814  &  $-$0&210167\\
\ce{NO+} &  $-$0&274894  &  $-$0&024556  &  $-$0&250338  &  $-$0&012280  &  $-$0&072516  &  $-$0&310580\\
\ce{LiH} &  $-$0&046844  &  0&000216  &  $-$0&047060  &  $-$0&000042  &  0&001223  &  $-$0&046094\\
\ce{HC#CH} &  $-$0&173054  &  0&012016  &  $-$0&185070  &  $-$0&000691  &  $-$0&068659  &  $-$0&266437\\
\ce{H2O\bond{...}H2O} &  $-$0&581202  &  $-$0&133722  &  $-$0&447480  &  0&003930  &  $-$0&100263  &  $-$0&410088\\
\hline
\end{tabular}
\end{center}
\label{tab:VeeTotales}
\end{table}
\thispagestyle{empty}
\thispagestyle{empty}

Before considering the splitting of the electronic energy in
accordance with the IQA method, we address the changes in (\textit{i}) the
spin components
$\Delta V_{\mathrm{ee}}^{\sigma \sigma}$ and $\Delta V_{\mathrm{ee}}^{\sigma \tau}$
with $\sigma \neq \tau$ and (\textit{ii}) the
exchange and correlation contributions of $V_{\mathrm{ee}}$
electron-electron repulsion for the complete system
 as reported in Table \ref{tab:VeeTotales}. 
As expected, the most important contribution to 
$\Delta V_{\mathrm{ee}}$ comes from the unlike-spin component $\Delta
V_{\mathrm{ee}}^{\sigma \tau}$, \ie $| \Delta V_{\mathrm{ee}}^{\sigma
\sigma} | < | V_{\mathrm{ee}}^{\sigma \tau} |$ because of the complete lack of
correlation for electrons with different spin projections (\ie
Coulomb correlation) in the
HF approximation.~\cite{McWeeny1992}
This condition holds even when $V_{\mathrm{ee}}^{\sigma \tau}$ and
$V_{\mathrm{ee}}^{\sigma \sigma}$ are weighted by the number of
electron pairs with the same and different spin projections, \ie
$N_{\sigma\sigma} = N_\sigma(N_\sigma - 1) + N_\tau(N_\tau - 1)$ and 
$N_{\sigma\tau} = 2N_\sigma N_\tau$
respectively. By con\-sid\-er\-ing the total number of electron
pairs,
$N_{\sigma\sigma,\sigma\tau} = N_{\sigma\sigma} + N_{\sigma\tau}$,
we note that the ratio $\Delta V_{\ee}/N_{\sigma\sigma,\sigma\tau}$
is in the range $1.0$-$4.0 \times 10^{-3}$ a.u. for all the considered
species. The absolute values $|\Delta
V_{\ee}/N_{\sigma\sigma,\sigma\tau}|$ are
greater for the ionic species, \eg\, LiF and LiH, 
than they are for the covalent molecules studied in this work such as
\ce{HC#CH}. Something similar occurs for the ratios $\Delta
V_{\ee}^{\sigma\sigma}/N_{\sigma\sigma}$ and $\Delta
V_{\ee}^{\sigma\tau}/N_{\sigma\tau}$ whose magnitudes are slightly smaller and
larger respectively than that of
$\Delta V_{\ee}/N_{\sigma\sigma,\sigma\tau}$. 
  
There are four systems
(\ce{BeH2}, \ce{BH}, \ce{LiH} and \ce{HC#CH}) for which $\Delta
V_{\mathrm{ee}}^{\sigma \sigma} > 0$ on account of
small positive changes of the same spin contributions to the coulombic part to
$V_{\mathrm{ee}}$.
We also note that apart from \ce{NO+}, the exchange component 
does not change substantially after the inclusion of electron
correlation and the reduction of the magnitude of $|V_{\mathrm{ee}}|$
occurs mainly through the correlation parts $\sigma \sigma$ and $\sigma
\tau$ ($\sigma \neq \tau$). 
The electron correlation component to
$V_{\mathrm{ee}}$ has a larger contribution from 
the unlike-spin electron pairs $V_{\mathrm{corr}}^{\sigma \tau}$
than for the like-spin pairs, %$\alpha \alpha$ and $\beta \beta$ pairs,
again in consistency with the previously mentioned absence of Coulomb correlation
in the HF method.
%stronger effect on $V_{\mathrm{ee}}^{\sigma \tau}$ that the
%corresponding on $V_{\mathrm{ee}}^{\sigma \sigma}$, because of the
%complete lack of correlation for electrons with different spin
%components at the Hartree-Fock level. 
%To analize further the change in correlation for the same and unlike
%spin components, we considered 
The ratio 
${V_{\mathrm{corr}}^{\sigma
\tau}}/{V_{\mathrm{corr}}^{\sigma \sigma}}$
%
%\vspace*{0.3cm}
%
%\begin{equation}
%\vartheta_{\mathrm{corr}} = 
%\end{equation}
%
%\vspace*{0.5cm}
%
%\noindent in which $\Delta V_{\mathrm{ee}}^{\mu \nu}$ is the change
%of $V_{\mathrm{ee}}^{\mu \nu}$ after electron correlation is taken
%into account. 
%which 
is around $3.5$--$6.0$ in most
of the considered molecules but it can be as large as $\approx$ $40$--$50$ in
magnitude, \eg, in \ce{LiH} and \ce{BeH2}. The
consideration of the $N_{\sigma\sigma}$ and $N_{\sigma\tau}$ pairs
does not change substantially the proportion
${V_{\mathrm{corr}}^{\sigma \tau}}/{V_{\mathrm{corr}}^{\sigma
\sigma}}$. This behaviour is expected, especially when one considers
that $N_{\sigma\tau}/N_{\sigma\sigma} \to 1$ when the number of
electrons increases.  
 We see thus how the consideration of the
spin-dependent matrices yields insights about the changes in Fermi
and Coulomb correlation due to the consideration of post-Hartree-Fock
methods, like coupled cluster theory in this case.

%Concerning the open shell species, 
%we also observe that $V_{\mathrm{ee}}^{\sigma \tau} >
%V_{\mathrm{ee}}^{\sigma \sigma}$ as in the closed shell systems. 

%We note that $\lambda_{\mathrm{ee}}$ is considerably
%larger for covalently bound molecules \eg \ce{CN-} and \ce{HCCH} for
%which $\lambda_{\mathrm{ee}} \approx 10$ than for ionic species such as \ce{LiF}
%with an associated $\lambda_{{ee}} \approx 3$. The weighting of
%$\Delta V_{\mathrm{ee}}^{\sigma \tau}$
%and $\Delta V_{\mathrm{ee}}^{\sigma
%\sigma}$ with the number of $\sigma \sigma$  
%($2 N_\sigma(N_\sigma - 1)$) 
%and 
%$\sigma \tau$ 
%%($2 N_\sigma N_\tau$)
%electron pairs (the number of these pairs being equal to $2
%N_\sigma(N_\sigma - 1)$ and $2 N_\sigma N_\tau$ respectively for closed shell
%systems) does not substantially change the comparison between closed shell covalent and
%ionic systems.  This means that the consideration of post Hartree-Fock
%methods covers a wide range for the observed changes in spin dependent
%correlation and this depends strongly on the chemical nature of the
%system under consideration.

\begin{table}

\caption{Changes in the $V_{\mathrm{ee}}$ component of the net
IQA energies along with its spin components $\Delta
V_{\mathrm{ee}}^{\sigma \sigma}$ and $\Delta V_{\mathrm{ee}}^{\sigma
\tau}$ ($\sigma \neq \tau$)  after the inclusion of dynamical electron
correlation. The change in the total exchange-correlation, along with
its spin components are shown as well. The first row for every system
correspond to the atom with the smallest atomic number. 
We averaged the quantities corresponding to the oxygen and hydrogen
atoms in \ce{H2O\bond{...}H2O}. 
The data are reported in Hartrees.}

\begin{center}
%\hspace*{-2.0cm}
\begin{tabular}{l|r@{.}l|r@{.}l|r@{.}l|r@{.}l|r@{.}l|r@{.}l}
\hline
System & \multicolumn{2}{c|}{$ \Delta V_{\mathrm{ee}}^{\mathrm{A}}$} & 
\multicolumn{2}{c|}{$\Delta V_{\mathrm{ee}}^{\mathrm{A} \sigma \sigma}$} &
\multicolumn{2}{c|}{$\Delta V_{\mathrm{ee}}^{\mathrm{A} \sigma \tau}$} & 
\multicolumn{2}{c|}{$\Delta V_{\mathrm{XC}}^{\mathrm{A}}$} & 
\multicolumn{2}{c|}{$\Delta V_{\mathrm{XC}}^{\mathrm{A} \sigma \sigma}$} &
\multicolumn{2}{c}{$V_{\mathrm{corr}}^{\mathrm{A} \sigma \tau}$} \\
\hline
\ce{BeH2} &  $-$0&030747  &  0&000011  &  $-$0&030757  &  $-$0&027455  & 0&001657  &  $-$0&029111\\
          &  $-$0&009627  &  0&005036  &  $-$0&014664  &  $-$0&021491  & $-$0&000896  &  $-$0&020596\\
\ce{BH} &  $-$0&031281  &  $-$0&000098  &  $-$0&031180  &  $-$0&026357  & 0&002364  &  $-$0&028718\\
        &  $-$0&037238  &  0&013560  &  $-$0&050798  &  $-$0&078967  & $-$0&007305  &  $-$0&071663\\
\ce{CN-} &  $-$0&008088  &  0&036952  &  $-$0&045040  &  $-$0&089627  & $-$0&013498  &  $-$0&076129\\
           &  $-$0&329992  &  $-$0&088812  &  $-$0&241180  &  $-$0&297322  & $-$0&054178  &  $-$0&243144\\
\ce{HF} &  $-$0&004332  &  $-$0&000126  &  $-$0&004204  &  $-$0&007546  & $-$0&001733  &  $-$0&005812\\
        &  $-$0&364841  &  $-$0&104492  &  $-$0&260346  &  $-$0&226815  & $-$0&035480  &  $-$0&191333\\
\ce{LiF} &  $-$0&009196  &  0&003474  &  $-$0&012670  &  $-$0&015166  & 0&000489  &  $-$0&015656\\
         &  $-$0&417846  &  $-$0&126050  &  $-$0&291798  &  $-$0&223119  & $-$0&028688  &  $-$0&194434\\
\ce{NO+} &  0&207927  &  0&143872  &  0&064048  &  $-$0&228209  & $-$0&074196  &  $-$0&154020\\
         &  $-$0&669512  &  $-$0&229762  &  $-$0&439752  &  $-$0&224439  & $-$0&007224  &  $-$0&217215\\
\ce{LiH} &  $-$0&031192  &  $-$0&000034  &  $-$0&031160  &  $-$0&028178  & 0&001474  &  $-$0&029652\\
         &  $-$0&013338  &  0&001390  &  $-$0&014730  &  $-$0&014926  & 0&000596  &  $-$0&015524\\
\ce{HC#CH} &  $-$0&014759  &  $-$0&000409  &  $-$0&014350  &  $-$0&015249  & $-$0&000653  &  $-$0&014596\\
           &  $-$0&180195  &  $-$0&024855  &  $-$0&155340  &  $-$0&251784  & $-$0&060650  &  $-$0&191134\\
\ce{H2O\bond{...}H2O} &  $-$0&402531 & $-$0&118225 & $-$0&284305 & $-$0&295465 & $-$0&064693 & $-$0&230772\\
                      & $-$0&005249 & $-$0&000072 & $-$0&005176 & $-$0&008228 & $-$0&001562 & $-$0&006666\\
%\ce{H2O\bond{...}H2O} &  $-$0&157580 &    0&039874 & $-$0&197454 & $-$0&201762 & $-$0&086048 & $-$0&115714\\ 
                      %& $-$0&013004 & $-$0&000749 & $-$0&012255 & $-$0&020168 & $-$0&015837 & $-$0&004331\\
                      	
\hline
\end{tabular}
\end{center}
\label{tab:VeeIntraCS}
\end{table}
\thispagestyle{empty}
\thispagestyle{empty}

Concerning the IQA partition, Tables \ref{tab:VeeIntraCS}  and
\ref{tab:VeeInter}
%and
%\ref{tab:VeeIntraRads} 
show respectively the electron-electron component of the 
IQA net and interaction energy (equation (\ref{particion})) of the species
considered in this study.  Since the inclusion of DC is reflected
mostly in the correlation rather than in the exchange part of $V_{\mathrm{ee}}$
as reflected in the analysis of the data in Table \ref{tab:VeeTotales}, we consider
together the exchange and correlation components of $V_{\mathrm{ee}}^{\sigma
\sigma}$   %\cc{Edu: no es simplemente $V_{\mathrm{ee}}$?}
through our analysis of the IQA net and interaction energies.
The comparison of the $\Delta V_{\mathrm{ee}}^{\mathrm{A}}$ and
$\Delta V_{\mathrm{ee}}^{\mathrm{AB}}$ data reveals that the
change in the terms corresponding to the IQA net energy represents most of the 90\% of the
reduction in electron-electron repulsion in all of the studied
systems. 
In fact,
there are some cases (\ce{CN-}, \ce{HF}, \ce{LiF} and most
conspicuously \ce{NO+}) for which the change in the
electron-electron repulsion for the atomic basins surpasses that of
the molecular species. %leading to positive values of $\Delta
%V_{\mathrm{ee}}^{\mathrm{AB}} $. 
This means that the inclusion of dynamical
correlation may lead to a considerable reduction of the intrabasin
electron-electron repulsion, $V_{\mathrm{ee}}^{\mathrm{A}}$, at the
expense of a considerable increase of this quantity for the
interatomic interaction energy. %as shown in Table \ref{tab:VeeInter}:
%most of the entries \color{red}{(O SINO HALF OF THE ENTRIES:
%DEPENDIENDO DE LO QUE RESULT\'E DEL D\'IMERO DE AGUA) }\color{black} of
%$\Delta V_{\mathrm{ee}}^{\mathrm{AB}}$ are greater than 0.  
This observation is consistent with previous descriptions of the
inclusion of electron correlation in chemical bonding~\cite{lewisAndBeyond}.
In agreement with the larger change in the Coulomb over the Fermi
correlation in the molecular electron-electron repulsion
(Table \ref{tab:VeeTotales}), the intra-atomic spin-dependent
electron-electron repulsion fulfil the conditions
%
%\vspace*{-0.5cm}
%
\begin{align}
 \Delta V_{\mathrm{ee}}^{\mathrm{A} \sigma \tau} - \Delta
V_{\mathrm{ee}}^{\mathrm{A} \sigma \sigma} & < 0, \label{sigmaTauSobreSigmaSigma_1} \\
 V_{\mathrm{corr}}^{\mathrm{A} \sigma \tau} - \Delta
V_{\mathrm{XC}}^{\mathrm{A} \sigma \sigma} & < 0, \label{sigmaTauSobreSigmaSigma_2}
\end{align}
the differences being in the interval of tens and even
hundreds of milliHartrees. That is to say, the magnitude of the change of the 
intra-atomic unlike-spin electron-electron repulsion, $\Delta
V_{\mathrm{ee}}^{\mathrm{A} \sigma
\tau}$, exceeds the corresponding value
for the same spin quantity, $\Delta V_{\mathrm{ee}}^{\mathrm{A} \sigma
\sigma}$. Since the change $\Delta V_{\mathrm{ee}}^{\mathrm{A} \sigma
\tau}$ is reflected through modifications of the Coulomb correlation then
the magnitude of $| V_{\mathrm{corr}}^{\mathrm{A} \sigma \tau}|$
exceeds that of $|V_{\mathrm{XC}}^{\mathrm{A} \sigma \sigma}|$ as
specified
in condition (\ref{sigmaTauSobreSigmaSigma_2}). Additionally,
the intra-atomic Coulomb correlation energies
($V_{\mathrm{corr}}^{\mathrm{A} \sigma \tau}$)
%reported in the last column of Table \ref{tab:VeeIntraCS} constitute
constitute indeed an important fraction of the molecular $\sigma\tau$
correlation as it can be appreciated by comparing 
the last columns of Tables \ref{tab:VeeTotales} and
\ref{tab:VeeIntraCS}.

The effect of the consideration of CC theory on the spin-dependent terms of the
IQA interaction energy is different to that 
of the IQA
net energy components.
For example and as discussed above, most of the entries of
$\Delta V_{\mathrm{ee}}^{\mathrm{AB}}$ in Table \ref{tab:VeeInter} 
indicate a slightly larger electron-electron repulsion among the
QTAIM basins on account of DC. In addition, the changes in the IQA spin-dependent
electron-electron repulsion terms, $\Delta V_{\mathrm{ee}}^{\mathrm{AB}
\sigma \sigma}$ and $\Delta V_{\mathrm{ee}}^{\mathrm{AB} \sigma \tau}$
on one hand along with $\Delta V_{\mathrm{XC}}^{\mathrm{AB} \sigma
\sigma}$ and $V_{\mathrm{corr}}^{\mathrm{AB} \sigma \tau}$ on the other,
do not meet conditions (\ref{sigmaTauSobreSigmaSigma_1}) and
(\ref{sigmaTauSobreSigmaSigma_2}). 
The change in the interatomic same-spin exchange-correlation, $\Delta
V_{\mathrm{XC}}^{\mathrm{AB}
\sigma \sigma}$ is, indeed, in most cases more neg\-a\-tive than
$V_{\mathrm{corr}}^{\mathrm{AB} \sigma \tau}$ (last two columns of
Table \ref{tab:VeeInter}).
In other words, the Fermi and
Coulomb correlation effects act differently on the IQA net and
interatomic energies: Coulomb correlation being overwhelmingly dominant in the
changes of $E_{\mathrm{net}}^{\mathrm{A}}$ energies, while Fermi correlation is
moderately more important in the change of
$E_{\mathrm{int}}^{\mathrm{AB}}$.
%EL ASUNTO DE QUE LA CORRELACIÓN TOTAL ES
%PARTICULARMENTE INTRATÓMICA.
%\begin{equation}
%\frac{\Delta V_{\mathrm{corr}}^{\mathrm{A} \sigma \tau}}
%{\Delta V_{\mathrm{corr}}^{\sigma \tau}} >
%\frac{\Delta V_{\mathrm{XC}}^{\mathrm{A}}}{\Delta V_{\mathrm{XC}}}
%\label{radiosIntra}
%\end{equation}
%\vspace*{0.5cm}
%
%\noindent wherein $\Delta X$ is the difference in quantity $X$ after
%considering the dynamic correlaton by means of HF/CC transition
%densities. The condition (\ref{radiosIntra}) implies that the
%intra-atomic changes on account of the consideration of DC contribute
%more to the unlike spin correlation than to the XC component.
%\color{red} Escribir algo acerca de las energ\'ias at\'omicas de los
%radicales \color{black}

%\color{red} Escribir sobre las energ\'ias de interacci\'on IQA.
%\color{black}

%The effect of DC diminishes the electron-electron respulsion in
%the IQA interatomic energies with the exception of \ce{NO+}.

%We had already pointed out in the analysis of Table \ref{tab:VeeIntraCS} that
%the exchange-correlation for the intra-atomic contributions of
%\ce{NO+} was larger than that observed for the whole molecule.\newline
%Cambios en el componente de intercambio y correlación de los sistemas
%Totales y después intraatómicos
%intra-atomic energies of the systems addresses in 

\begin{table}

\caption{Differences in the $V_{\mathrm{ee}}$ interaction IQA energies
related to covalent and H-bond in \ce{H2O\bond{...}H2O} and its
spin-dependent contributions $\Delta V_{\mathrm{ee}}^{\sigma
\sigma}$ and $\Delta V_{\mathrm{ee}}^{\sigma \tau}$ on account of the
consideration of electron correlation by means of HF/CC transition
densities. The changes in the total exchange correlation energies
along with its same and unlike spin contributions are reported too.
The first and second entries for \ce{HC#CH} are the \ce{H-C} and
\ce{C#C} bonds respectively, while those for
\ce{H2O\bond{...}H2O} are the H-bond and the \ce{O-H} covalent
linkage. Atomic units are used throughout.}

\begin{center}
%\hspace*{-2.0cm}
\begin{tabular}{l|r@{.}l|r@{.}l|r@{.}l|r@{.}l|r@{.}l|r@{.}l}
\hline
System & \multicolumn{2}{c|}{$ \Delta V_{\mathrm{ee}}^{\mathrm{AB}}$} & 
\multicolumn{2}{c|}{$\Delta V_{\mathrm{ee}}^{\mathrm{AB} \sigma
\sigma}$} &
\multicolumn{2}{c|}{$\Delta V_{\mathrm{ee}}^{\mathrm{AB} \sigma \tau}$}
& 
\multicolumn{2}{c|}{$\Delta V_{\mathrm{XC}}^{\mathrm{AB}}$} & 
\multicolumn{2}{c|}{$\Delta V_{\mathrm{XC}}^{\mathrm{AB} \sigma \sigma}$} &
\multicolumn{2}{c}{$V_{\mathrm{corr}}^{\mathrm{AB} \sigma \tau}$} \\
\hline
\ce{BeH2} &  $-$0&001851  &  $-$0&000721  &  $-$0&001131  &  $-$0&003913  & $-$0&001753  &  $-$0&002163 \\
\ce{BH} &  $-$0&007729  &  $-$0&004212  &  $-$0&003516  &  $-$0&007219  & $-$0&003957  &  $-$0&003261 \\
\ce{CN-} &  0&103162  &  0&035114  &  0&068048  &  0&048460  & $-$0&001079  &  0&049539 \\
\ce{HF} &  0&044980  &  0&019462  &  0&025518  &  $-$0&003934  & $-$0&004994  &  0&001061 \\
\ce{LiF} &  0&008182  &  0&003150  &  0&005030  &  $-$0&002034  & $-$0&001956  &  $-$0&000077 \\
\ce{NO+} &  0&186700  &  0&061334  &  0&125366  &  0&042729  & $-$0&003588  &  0&046313 \\
\ce{LiH} &  $-$0&002309  &  $-$0&001140  &  $-$0&001170  &  $-$0&001807  & $-$0&000889  &  $-$0&000918 \\
\ce{HC#CH} &  0&049576  &  0&017434  &  0&032142  &  0&046241  & 0&015767  &  0&030474 \\
           &  0&072880  &  0&004180  &  0&068700  &  0&062596  & $-$0&000962  &  0&063558 \\
\ce{H2O\bond{...}H2O} &  0&052647 & 0&025984 & 0&026664 & 0&017908 & 0&008614 & 0&009294 \\
                      &  0&061752 & 0&026097 & 0&035655 & 0&030718 & 0&010580 & 0&020138 \\
%\ce{H2O\bond{...}H2O} &  0&020861 & 0&010330 & 0&010532 & 0&065032 & 0&032618 & 0&032415 \\
                     % &  0&085414 & 0&039126 & 0&046288 & 0&031173 & 0&015944 & 0&015228 \\
\hline
\end{tabular}
\end{center}
\label{tab:VeeInter}
\end{table}
\thispagestyle{empty}
\thispagestyle{empty}

%%% Resultados de DIs. Quiza el enfoque tiene que ser distinto pero
% primero hay que discutir la parte de IQAs... dejo la discusion aqui
% como si no hubiera IQA por si inspira la discusion de las mismas...

Since the exchange-correlation of the IQA interaction energy is
related with the QTAIM delocalisation
indices,~\cite{bondPathsPrivileged} we consider now the separate
Fermi and Coulomb correlation effects in the DIs.
Table~\ref{tablaDIs} collects the LI and DI values for the series of
molecules studied. The HF/CC LI and DI are in reasonable agreement with
the CISD/6-311++G(2d,2p) results published in
Ref.~[\!\!\citenum{matito:07fd}]
for the series of molecules studied in both papers (\ce{CN-}, \ce{HF},
\ce{LiF}, \ce{NO+} and \ce{LiH}),
indicating that ($i$) the present CC calculations introduce a similar amount
of DC and ($ii$) the electron correlation is sufficiently well described by
the HF/CC pair density. 
Unlike the CISD results, the approximate DI values calculated from
M\"uller's approximation of the pair density ($\delta^{\text{AB}}_F$) give a very
poor agreement with the HF/CC results, giving values which are actually closer to
the (uncorrelated) HF values.  The same occurs for the 
HF-like ($\delta^{\text{AB}}_A$) approximation. Therefore, we 
conclude that the HF/CC first-order reduced density matrices give a very
deficient approximation of electron correlation effects. 
Despite second-order HF/CC matrices reduce to first-order
HF/CC ones (see Equations (\ref{rho1Rho2Int}) and (\ref{rho1Rho2Sum})), the second-order HF/CC matrices
provide reasonably accurate DIs while first-order HF/CC matrices used
on DI approximations (which usually provide sensible
results~\cite{wang:03jcc,matito:07fd,dmftiqa,feixas:10jctcelf})
do not improve HF results.  

Upon separation of the DI into spin components, we observe that
Fermi's correlation is reasonably well reproduced by the HF-like
approximation, as one can infer by the small differences between 
$\delta^{\text{AB}}_A$ and $\delta^{\text{AB},\sigma\sigma}$. 
The comparison with CISD values~\cite{matito:07fd} reveals that Fermi's correlation is
quite well reproduced by the HF/CC like-spin pair density expressions.
The role of the Coulomb correlation is more obvious for those molecules
that present a strong covalent bond, such as \ce{CN-} and \ce{NO+}~\cite{matito:07fd}.
The $\delta^{\text{AB},\sigma\tau}$ ($\sigma \neq \tau$) values are indeed larger for these
species, however, not as large as the values reported for the CISD wavefunction
($\delta^{\mathrm{C,N}}_{\sigma\tau,\,\mathrm{CISD}}=-0.379$ and 
$\delta^{\mathrm{N,O}}_{\sigma\tau,\,\mathrm{CISD}}=-0.538$). These numbers put forward that
the HF/CC cross-spin pair density expressions underestimate Coulomb correlation
to some extent. %It is expected that the Lagrangian CC pair density values can
%improve this behavior.~\cite{ccLag} 
Overall, we can safely conclude that CC/HF pair density
expressions are adequate to describe ionic and weak-interaction molecules but
underestimate the Coulomb correlation effects in covalent bonds, leading to
an overestimation of DI.

 A better 
consideration of DC in delocalisation indices by means of coupled
cluster theory warrants further investigation in approximated CC
density matrices. 

%\cc{It is expected that Lagragian CC density matrices correct
%this behavior. We are currently investigating this issue in our laboratories.}

\thispagestyle{empty}
\thispagestyle{empty}
\cc{
\begin{table}
\caption{DIs using HF/CC density matrices ($\delta^{\text{AB}}$) and their
decomposition into spin cases according to Eq.~\ref{DIdecomp}
($\delta^{\text{AB},\sigma\sigma}$ and $\delta^{\text{AB},\sigma\tau}$).
DIs from Hartree-Fock-like approximation (Eq.~\ref{HFL}) of the pair density ($\delta_{A}^{\text{AB}}$), 
from M\"uller's approximation of the pair density ($\delta_F^{\text{AB}}$) and Hartree-Fock value
$\delta_{\text{HF}}^{\text{AB}}$. The same-atom values refer to localization indices (Eq.~\ref{lambda}).
}
\label{tablaDIs}
\begin{center}
%\begin{tabular}{l||r@{.}l|r@{.}l||r@{.}l|r@{.}l|r@{.}l}
\begin{tabular}{lccccccc}
\hline
 & $A-B$ & $\delta^{\text{AB}}$ & $\delta^{\text{AB},\sigma\sigma}$ & $\delta^{\text{AB},\sigma\tau}$ & $\delta_{A}^{\text{AB}}$ & $\delta_F^{\text{AB}}$ & $\delta_{\text{HF}}^{\text{AB}}$ \\
\hline
\ce{BeH2} & Be-Be & 2.035 & 2.025 & 0.009 & 2.023 & 2.022 & 2.021\\
 & Be-H & 0.331 & 0.340 & -0.010  & 0.342 & 0.343 & 0.335 \\
 & H-H' & 0.074 & 0.075 & -0.001 & 0.074 & 0.074 & 0.072 \\
 & H-H & 1.614 & 1.609 & 0.005 & 1.607 & 1.606 & 1.617 \\
\ce{BH} & B-B & 3.934 & 3.919 & 0.015 & 3.918 & 3.915 & 3.918 \\
 & B-H & 0.665 & 0.695 & -0.030 & 0.699 & 0.704 & 0.685 \\
 & H-H & 1.400 & 1.386 & 0.015 & 1.384 & 1.381 & 1.397 \\
\ce{CN-} & C-C & 4.426 & 4.288 & 0.138 & 4.236 & 4.224 & 4.154\\
 & C-N & 1.979 & 2.256 & -0.277 & 2.362 & 2.382 & 2.238\\
 & N-N & 7.591 & 7.452 & 0.139 & 7.401 & 7.389 & 7.609\\
\ce{HF} & H-H & 0.040 & 0.032 & 0.008 & 0.031 & 0.030 & 0.028\\
 & H-F & 0.450 & 0.467 & -0.017 & 0.469 & 0.471 & 0.450\\
 & F-F & 9.509 & 9.501 & 0.008 & 9.500 & 9.499 & 9.522\\
\ce{LiF} & Li-Li & 1.976 & 1.975 & 0.001 & 1.974 & 1.974 & 1.974\\
 & Li-F & 0.195 & 0.197 & -0.002 & 0.198 & 0.199 & 0.186\\
 & F-F & 9.829 & 9.828 & 0.001 & 9.830 & 9.827 & 9.839\\
\ce{NO+} & N-N & 4.559 & 4.399 & 0.160 & 4.340 & 4.321 & 4.288\\
 & N-O & 1.999 & 2.319 & -0.321 & 2.438 & 2.475 & 2.358\\
 & O-O & 7.443 & 7.282 & 0.160 & 7.224 & 7.205 & 7.354\\
\ce{LiH} & Li-Li & 1.995 & 1.994 & 0.002 & 1.994 & 1.993 & 1.993\\
 & Li-H & 0.218 & 0.221 & -0.003 & 0.222 & 0.222 & 0.215\\
 & H-H & 1.787 & 1.785 & 0.002 & 1.785 & 1.785 & 1.793\\
\ce{H2O\bond{...}H2O} & \ce{O\bond{...}H} & 0.061 & 0.060 & 0.001 & 0.060 & 0.060 & --- \\
\ce{HCCH} & C-C  & 4.571 & 4.293 &  0.278 & 4.225 & 4.223 & 4.223 \\
          & C-C' & 2.242 & 2.735 & -0.493 & 2.863 & 2.864 & 2.863 \\
          & C-H  & 0.884 & 0.956 & -0.072 & 0.959 & 0.960 & 0.961 \\
\hline
\end{tabular}
\end{center}
\end{table}
}
\thispagestyle{empty}
\thispagestyle{empty}

\section{Concluding remarks}

We have considered spin-dependent one- and two- electron matrices based
on HF and HF/CC transition densities to evaluate separately the Fermi
and Coulomb correlations consequences on the IQA electronic energy partition. The
results show that the net unlike-spin correlation is the dominant
factor in the reduction of the electron-electron repulsion across the
system to the extent that in some cases it surpasses the decrease of
$V_{\mathrm{ee}}$ in the whole molecule or molecular cluster. This
situation leads to an increase of the electronic repulsion among the QTAIM
basins. Overall, different Fermi and Coulomb correlations effects are
observed in the IQA net and interaction energies. The same
spin-dependent density matrices were used to determine the impact of
these two types of correlation in QTAIM delocalisation indices. Our results
show that although $\varrho_2^{\HF/\CC}(\mathbf{r}_1,\mathbf{r}_2)$ 
and $\varrho_1^{\HF/\CC}(\mathbf{r}_1;\mathbf{r}_1^{\, \prime})$
in conjuntion can give a proper account of electron correlation on
the DIs, care must be taken in
the consideration of approximations based only on
the latter scalar field. Altogether, we expect that the approach
presented in this work prove useful in the evaluation of Fermi and
Coulomb effects both in quantum chemical topology and physical chemistry.

\begin{acknowledgements}
%This research has been funded by the Spanish MINECO
E.F. and A.M.P. thank the Spanish MINECO Project CTQ2012-31174.
E.M. also expresses his gratitude to Spanish MINECO Project No.
CTQ2014-52525-P and to the Basque 
Country Consolidated Group Project No. IT588-13. T.R.R. acknowledges
financial support from
CONACyT/Mexico Project No. 253776 and PAPITT/UNAM Project IN209715
along with computer time from DGTIC/UNAM grant SC16-1-IG-99.
I.R. and F.J.H-G. are grateful to CONACyT/Mexico for
the Ph.D. scholarships 255243 and 288853. T.R.-R. is thankful to
Magdalena Aguilar Araiza, Gladys Cort\'es Romero and David
V{\'a}zquez Cuevas for technical support.
\end{acknowledgements}

% BibTeX users please use
\footnotesize{
\bibliographystyle{jcp}
\bibliography{referencias}   % name your BibTeX data base
}
%\nocite{*}

\end{document}